\documentclass[11pt]{article}

\newcommand{\LANLNumber} {0205038}

\usepackage[square,comma]{natbib}
\usepackage{epsfig}
\usepackage{amssymb}
\newcommand{\Order}     {\ensuremath{O}}
\newcommand{\alphas}    {\ensuremath{\alpha_\mathrm{s}}}

\newcommand{\GeV}       {\ensuremath{\mathrm{GeV}}}
\newcommand{\BB}        {\ensuremath{\rm B\overline{B}}}
\newcommand{\bulnu}     {\ensuremath{\rm b \to u\ell\nu}}
\newcommand{\bclnu}     {\ensuremath{\rm b \to c\ell\nu}}
\newcommand{\mSqCutHad} {\ensuremath{m^2_{\rm cut}}}
\newcommand{\mCutHad}   {\ensuremath{m_{\rm cut}}}
\newcommand{\mRefHad}   {\ensuremath{m_{\rm ref}}}
\newcommand{\shmax}     {\ensuremath{s_{\rm h}^{\rm max}}}

\newcommand{\mH}        {\ensuremath{m_{\rm h}}}
\newcommand{\mHSq}      {\ensuremath{m_{\rm h}^2}}
\newcommand{\EplusP}    {\ensuremath{E_{\rm W}+|\vec{p}_{\rm W}|}}
\newcommand{\EplusPave} {\ensuremath{\langle E_{\rm W}+|\vec{p}_{\rm W}|\rangle}}
\newcommand{\Vub}  {\ensuremath{|V_{ub}|}}
\newcommand{\epem} {\ensuremath{e^+e^-}}
\newcommand{\FourS}{\ensuremath{\Upsilon(4S)}}
\newcommand{\btosgam}   {\ensuremath{b \to s \gamma}}

\def\etal{et al.}

\begin{document}
{\thispagestyle{empty}

\begin{flushright}
hep-ex/\LANLNumber \\
June 2002 \\
\end{flushright}

\par\vskip 4cm

\begin{center}
  \Large\bf Complementary Observables for the Determination 
  of $\Vub$ in Inclusive Semileptonic B Decays
\end{center}
\bigskip

\begin{center}
  \large
  Robert V. Kowalewski$^{\rm a,1}$, Sven Menke$^{\rm b,2}$\\[0.2cm]
  $^{\rm a}${\em Department of Physics and Astronomy, University of Victoria,
  P.O. Box 3055, Victoria, BC V8W3P6, Canada}\\
  $^{\rm b}${\em Stanford Linear Accelerator Center,
    Stanford University, Stanford, CA 94309, U.S.A.}
\end{center}
\bigskip \bigskip

\begin{center}
\large \bf Abstract
\end{center}
 The determination of $\Vub$ from inclusive semileptonic B
  decays is limited by uncertainties in modeling the decay
  distributions in $\bulnu$ transitions.  The largest uncertainties
  arise from the limited knowledge of the appropriate b quark mass
  and Fermi momentum to use in the parameterization of the
  shape function.  This paper presents a new method in which
  these shape function parameters are constrained by the same
  data used to measure $\Vub$.  The method requires measurements
  of the momenta of both the charged lepton and the neutrino in
  semileptonic B decays.  From these
  quantities two complementary observables can be constructed,
  one for discriminating between $\bulnu$ transitions and background
  and the other for constraining the shape function.
  Using this technique the uncertaintites in $\Vub$ from
  the shape function may be signficantly reduced.  

\vfill
\begin{center}
To appear in Physics Letters {\bf B}
\end{center}

\vspace{1.0cm}
\begin{center}
{\em Stanford Linear Accelerator Center, Stanford University, 
Stanford, CA 94309} \\ \vspace{0.1cm}\hrule\vspace{0.1cm}
$^{\rm 1}$Work supported in part by the National Sciences and Engineering Research Council of Canada.\\
$^{\rm 2}$Work supported in part by Department of Energy contract DE-AC03-76SF00515.
\end{center}

\section{Introduction}

The precise determination of \Vub\ is of comparable importance to the
measurement of CP asymmetries in probing the CKM sector of the
Standard Model.  The most precise determination of \Vub\ at present
comes from a measurement of the endpoint of the charged lepton energy
spectrum in inclusive semileptonic ${\rm B}$ decays\cite{CLEO-vub},
which has a total relative uncertainty of 15\%.  The major
uncertainties in the endpoint measurement arise from the background
from $\bclnu$ decays and the theoretical prediction of the fraction of
the $\bulnu$ spectrum in the endpoint region.  These two sources of
error have opposite sensitivities to the lepton energy cut; raising
the cut decreases the former and increases the latter.  The total
error is minimized with a lepton energy cut of $\sim 2.2\, \GeV$.
Within the context of the lepton endpoint method, further progress
requires an improved understanding of the shape function
that describes the ${\rm b}$ quark mass and momentum distribution in
the ${\rm B}$ meson.

A number of proposals have been made to reduce the theoretical uncertainties
in the determination of \Vub.  These include measuring the recoil mass
of the hadronic system in a \bulnu\ transition\cite{recoilmass}, measuring
the invariant mass squared of the lepton pair\cite{qsquared} or using
a combination of variables\cite{combinedqsqmh}.  In addition, the use
of the photon energy spectrum in $\btosgam$ decays to reduce
shape function-related uncertainties
in the extraction of \Vub\ from semileptonic decays was proposed
in Ref.~\cite{bsgamIdea} and applied in Ref.~\cite{CLEO-vub}; 
more recent work\cite{bsgamNow} suggests that
the non-universality of the shape function may significantly limit
the precision achievable with this approach.

In this paper we discuss a technique by which the theoretical uncertainty
in extracting \Vub\ from inclusive semileptonic decays at the
${\rm \FourS}$ may be
significantly reduced.  The method requires the reconstruction of the
neutrino momentum vector in addition to that of the charged lepton in
semileptonic ${\rm B}$ decays.\footnote{Technical issues involving neutrino
reconstruction will not be discussed here; such
reconstruction has already been used in analyses of exclusive
semileptonic decays~\cite{CLEO-rholnu}.}  Using this information
two quantities are measured, one directly sensitive to the shape function
parameters and another offering good discrimination
between \bulnu\ and \bclnu\ decays.  This approach using two complementary
observables could be used in measurements of \Vub\ at 
present $\epem$ ${\rm B}$ factories.

\section{The $\bulnu$ Generator}

A Monte Carlo generator has been implemented to study
$\bulnu$ decays.  The generator 
is based on the triple differential decay width 
of Ref.~\cite{art:NeubertFazio},
which is calculated including ${\Order}(\alphas)$ corrections.
The parton-level distributions are convolved with a shape function
to obtain distributions in the experimental observables.  The shape 
function can be written in terms of a single variable 
$k_+$.  The b quark mass appearing in the parton-level distributions
is replaced by $m_{\rm b} + k_+$.  For the
distribution of $k_+$ we take~\cite{art:NeubertFazio}
\begin{equation}
 F(k_+) = N (1-\kappa)^a e^{(1+a)\kappa};\quad \kappa = \frac{k_+}{m_{\rm B}-m_{\rm b}} \le 1\ .
\label{eq:fermi_motion}
\end{equation}
The shape function can be parameterized using $m_{\rm b}$ and $a$ or,
equivalently, using $\overline{\Lambda}= m_{\rm B}-m_{\rm b}$ and
$\lambda_1 = -3\overline{\Lambda}\,{}^2/(1+a)$.

\section{Observables}

Detection of a charged lepton at an momentum
high enough to allow identification is the starting point for any
analysis of inclusive semileptonic decays.
The $\FourS$ decays into $\BB$, giving rise to ${\rm B}$ mesons that
are nearly at rest, with velocity 
$\beta = \sqrt{1-(2 m_{\rm B} / m_{\rm \FourS})^2} \simeq 0.06$.
The lepton momenta in the ${\rm B}$ meson rest frame are smeared by
the ${\rm B}$ motion in the $\FourS$ rest frame.  The
invariant masses of the lepton pair and of the recoiling hadronic
system are unaffected by the boost.

Two types of observables are constructed.
The first type allows discrimination between \bulnu\
decays and background.  For this purpose, the invariant mass of the
hadronic recoil, $\mH$, can be used, provided the other ${\rm
B}$ meson in the event is fully reconstructed.  Alternatively, a
measurement of the missing momentum, which can be obtained without
fully reconstructing the other ${\rm B}$ meson, can be combined with
the charged lepton momentum to determine $q^2$, the invariant
mass squared of the lepton pair.  
The measurement of $q^2$, in contrast to the neutrino
energy itself, does not suffer from the unknown direction of the parent
${\rm B}$ meson in the ${\rm \FourS}$ frame.  
\begin{figure}[htbp]
\begin{center}
 \resizebox{0.47\textwidth}{!}{%
    \includegraphics{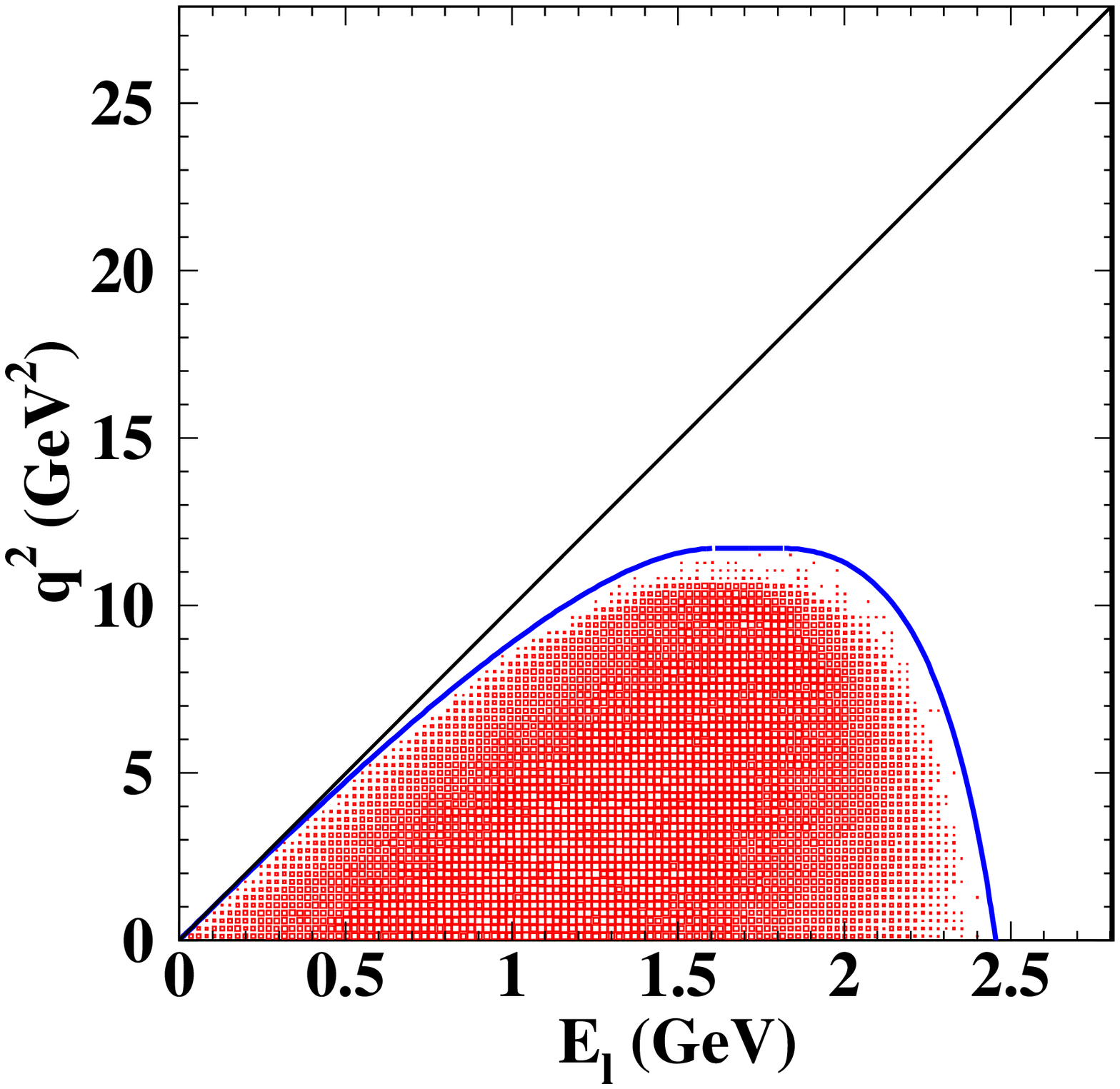}}
 \resizebox{0.47\textwidth}{!}{%
    \includegraphics{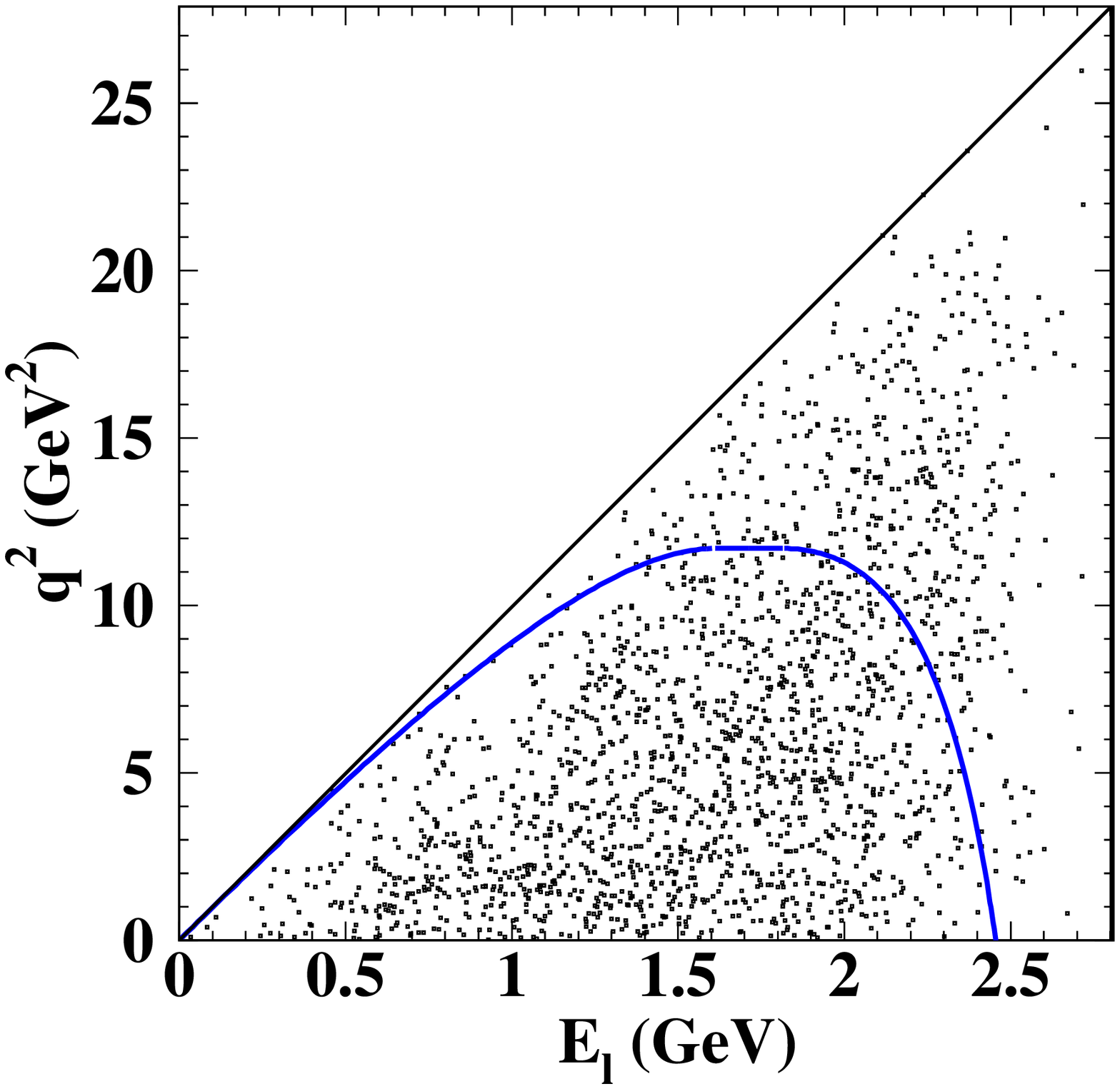}}
\end{center}
\begin{center}
\caption{$q^2$ vs. $E_\ell$ in the ${\rm \FourS}$ rest frame. The diagonal
line and the curve are contours for $\shmax=0$ and $m^2_{\rm D}$, 
respectively.  
The left plot shows \bclnu\ transitions and the right plot shows
\bulnu\ transitions. \label{fig-q2El}}
\end{center}
\end{figure}
Using the charged lepton energy and $q^2$ one
can define a variable to discriminate between $\bulnu$ and
$\bclnu$ decays.  In the B meson rest frame, ignoring lepton masses, 
the neutrino energy
satsifies $E_\nu \ge q^2/4E_\ell$.  Setting the neutrino energy equal to
its minimum value determines the maximum 
kinematically allowed invariant mass squared
of the hadronic recoil system:
\begin{equation}
\shmax = m_{\rm B}^2 + q^2 - 2 m_{\rm B} 
E_\ell\,  - 2 m_{\rm B} \left(\frac{q^2}{4E_\ell}\right) 
\quad\quad {\rm (B\ rest\ frame)}
\end{equation}
Accounting for the boost into the $\FourS$ frame and respecting
the condition $m_h + \sqrt{q^2} \le m_{\rm B}$ gives
\begin{eqnarray}
\shmax &=& m_{\rm B}^2 + q^2 - 2 m_{\rm B} 
E_\ell\, \sqrt{\frac{1\mp\beta}{1\pm\beta}}
 - 2 m_{\rm B} \left(\frac{q^2}{4E_\ell}\right) 
\sqrt{\frac{1\pm\beta}{1\mp\beta}} \nonumber \\
&& \quad {\rm for\ } \pm E_\ell>\pm \frac{1}{2}\left(m_{\rm B}-\mRefHad\right)
\sqrt{\frac{1\pm\beta}{1\mp\beta}}\ ,\\
\shmax &=& m_{\rm B}\, \mRefHad - 
\frac{
\mRefHad\, q^2 }{m_{\rm B} - \mRefHad} \quad {\rm otherwise.\nonumber }
\label{eqn-shmax-hi}
\end{eqnarray}
In the above $E_\ell$ is
the charged lepton energy in the ${\rm \FourS}$ rest frame.
An appropriate choice of
reference mass for the definition of $\shmax$ is $M_{\rm D}$.
Events containing $\bulnu$ decays are selected by requiring
$\shmax < \mSqCutHad$, with 
$\mCutHad\simeq m_{\rm D}$; no event with a true 
hadronic recoil mass above $\mCutHad$ can enter the sample unless 
$q^2$ (or, less likely, $E_\ell$) is misreconstructed.
Constant values of $\shmax$ map out contours in the $q^2$-$E_\ell$
plane as shown in Fig.~\ref{fig-q2El}.  
Decays from the dominant \bclnu\ process cannot contribute in the
region $\shmax<m^2_{\rm D}$ unless they are mismeasured.  

As seen in Fig.~\ref{fig-shmax}, the 
shape of the \bulnu\ spectrum in $\mHSq$ is sensitive to both
the assumed ${\rm b}$ quark mass and to the Fermi momentum.
The spectrum in $\shmax$ depends on $m_{\rm b}$ but 
is largely insensitive to
the Fermi momentum; given that it is based largely on $q^2$, this is
as expected.  In these and subsequent plots
the charged lepton energy is required to exceed $1\,\GeV$
in the ${\rm \FourS}$ frame to ensure that it can be cleanly
identified.\footnote{The efficiency of this requirement is 87\%\ and
is not very sensitive to the shape function.}
\begin{figure}[htbp]
\begin{center}
    \resizebox{0.49\textwidth}{!}{\includegraphics{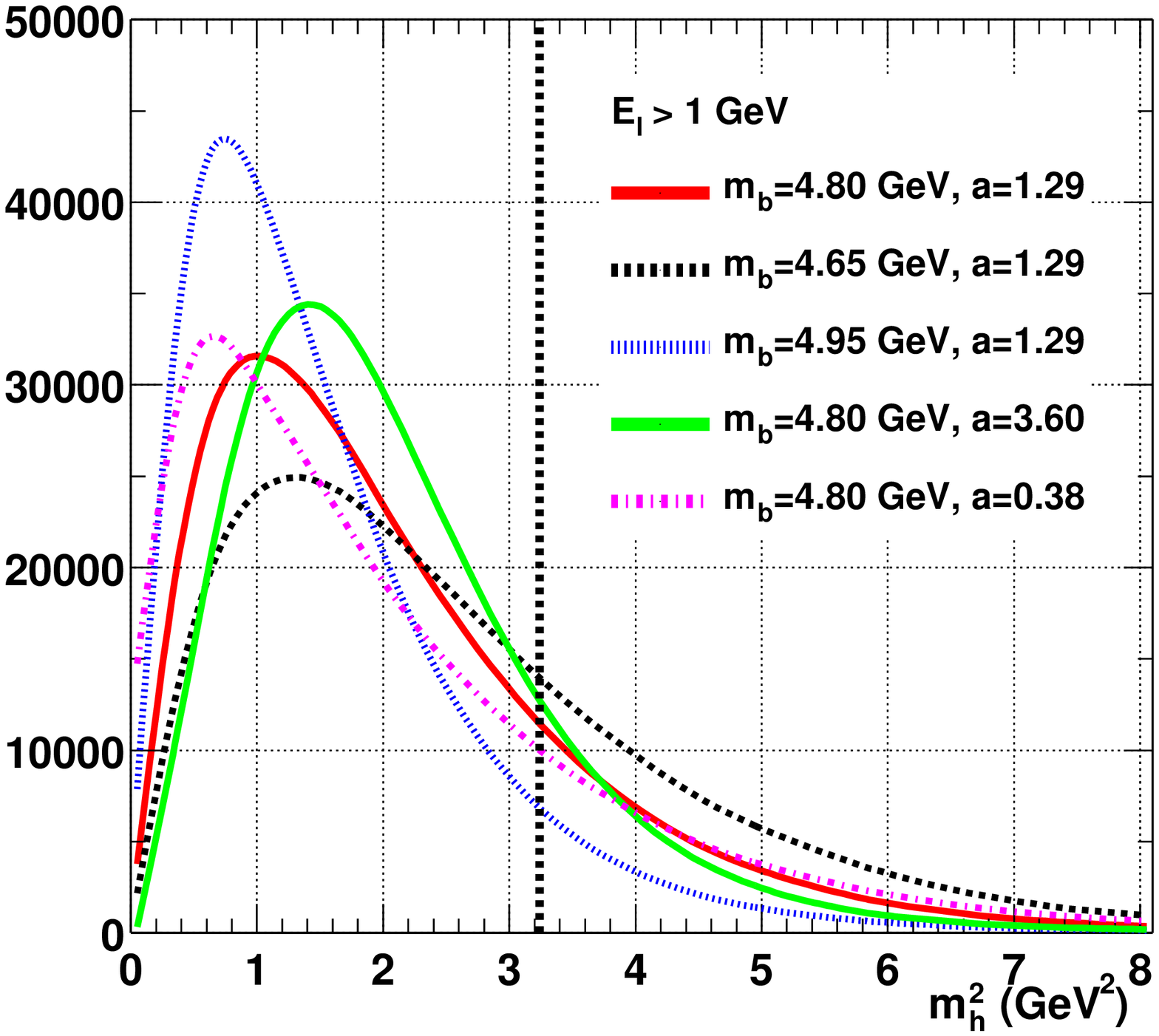}}
    \resizebox{0.49\textwidth}{!}{\includegraphics{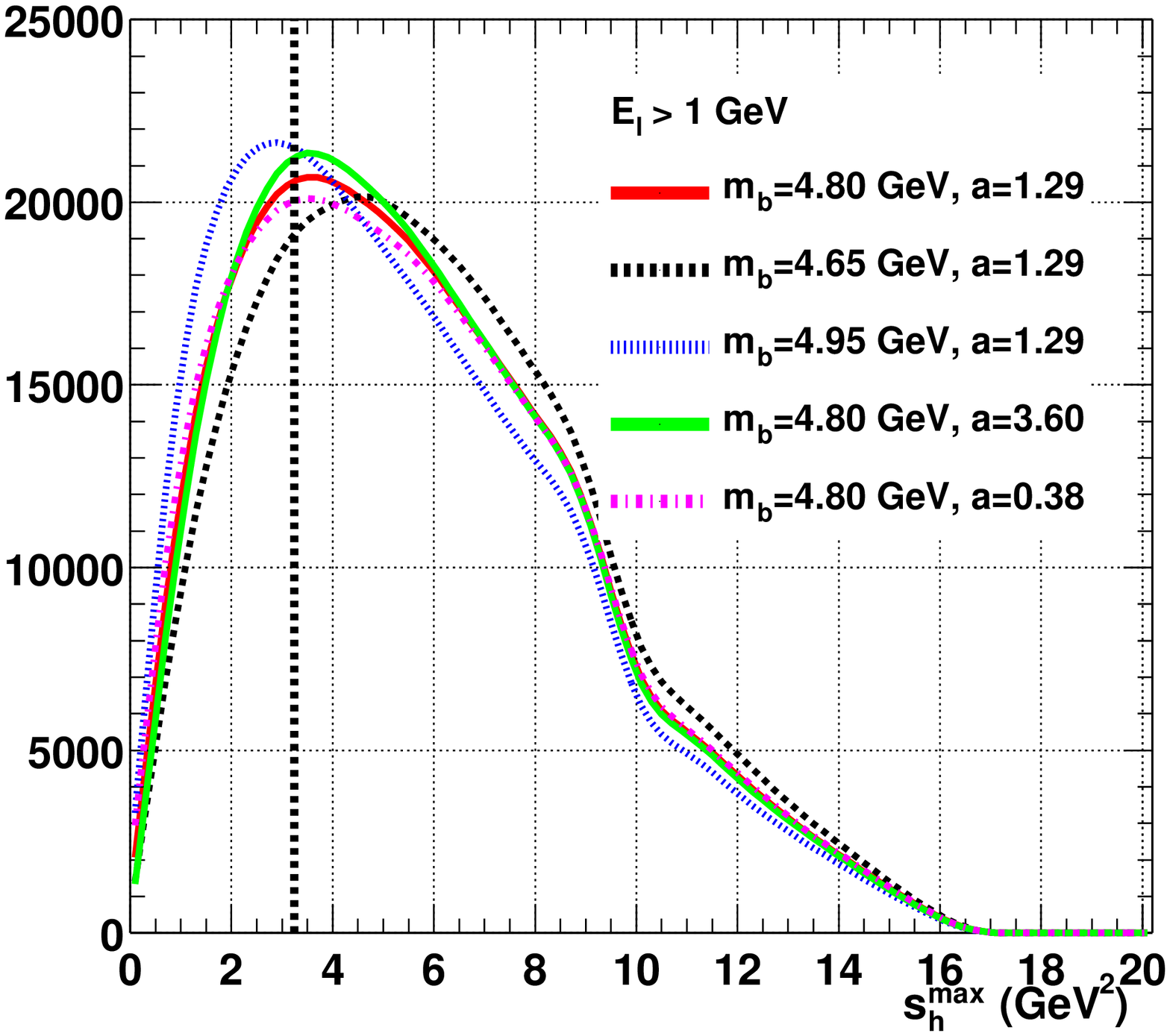}}
\end{center}
\begin{center}
\caption{The effect of varying the shape function parameters
on the distributions of $\mHSq$ and \shmax\ for \bulnu\ decays.
In each case the charged lepton
exceeds $1\,\GeV$ in the ${\rm \FourS}$ frame.  
The \bulnu\ signal region is to the
left of the dashed vertical line. \label{fig-shmax}}
\end{center}
\end{figure}

The second type of observable is sensitive to the effective ${\rm b}$
quark mass.  In the rest frame of the ${\rm b}$ quark, $m_{\rm b}$ is
related to the W properties:
$$
E^{\,({\rm b})}_{\rm W} = \frac{m_{\rm b}^2 + q^2 - m^2_{\rm u}}{2 m_{\rm b}}
\quad\Longrightarrow
\quad m_{\rm b} \simeq E^{\,({\rm b})}_{\rm W}+|\vec{p}^{\,({\rm b})}_{\rm W}|
$$
where $m_{\rm u}$ is the
effective mass of the final state u quark
and the equality in the last relation holds in the limit
${m_{\rm u}\ll m_{\rm b}}$.
The quantity $E^{\,({\rm b})}_{\rm W} + |\vec{p}^{\,({\rm
b})}_{\rm W}|$ is 
an estimator for the effective ${\rm b}$
quark mass.  The boost from the ${\rm b}$ quark rest frame into the
${\rm \FourS}$ rest frame smears the estimate but introduces little
bias, since the boost direction is uncorrelated with 
the W direction
and $\gamma\simeq 1$.  The average value
$\EplusPave$, measured in the ${\rm \FourS}$ rest frame from the
lepton momenta, is an observable with substantial 
sensitivity to the effective ${\rm b}$ quark mass.

The strategy for determining \Vub\ uses either 
\shmax\ or $\mH$ to separate
\bulnu\ decays from background and \EplusPave\ to limit the
variation in shape function parameters that must be considered.  

How well can $\EplusPave$ be measured?  The r.m.s. of the generated
$\EplusP$ distribution is $\sim 0.25\,\GeV$ after cuts are applied on
$\shmax$ or $\mH$ to select \bulnu\ events.  The experimental
resolution on missing energy will further broaden this; the r.m.s.
missing energy resolution in the $\epem$ B factories is ${\Order}(0.2\,\GeV)$.
With B-factory data samples the statistical uncertainty in 
$\EplusPave$ can be reduced below $\sim 0.02\,\GeV$.  However, it will 
be an experimental challenge to understand 
systematic biases\footnote{Biases not properly
accounted for in Monte Carlo simulations.  Note, however, that
uncertainties on the missing energy {\em resolution} do
not affect $\EplusPave$ directly.} at this level.

\section{Extracting $\Vub$}

\begin{figure}[htbp]
\begin{center}
 \resizebox{0.49\textwidth}{!}{\includegraphics{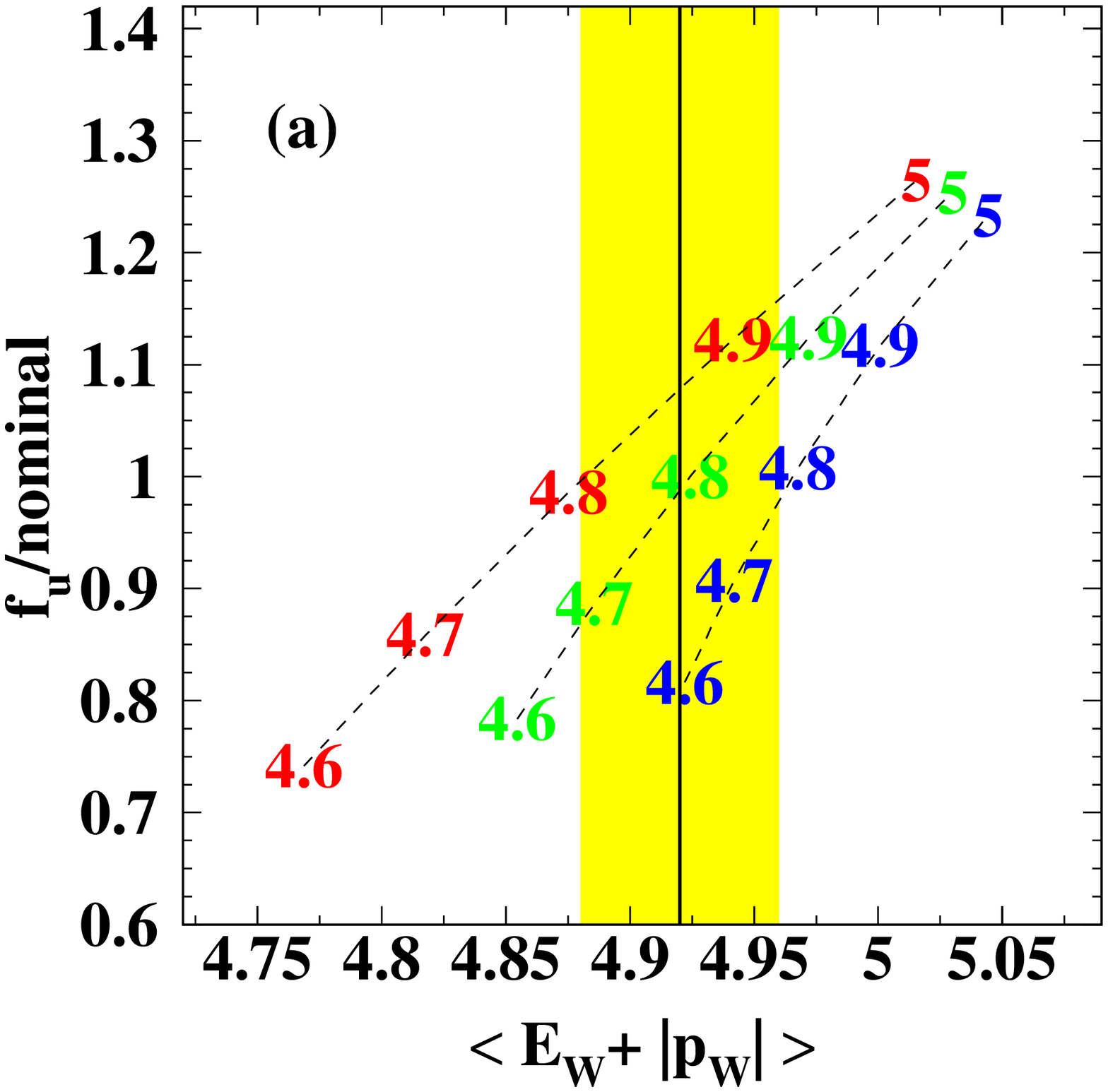}}
 \resizebox{0.49\textwidth}{!}{\includegraphics{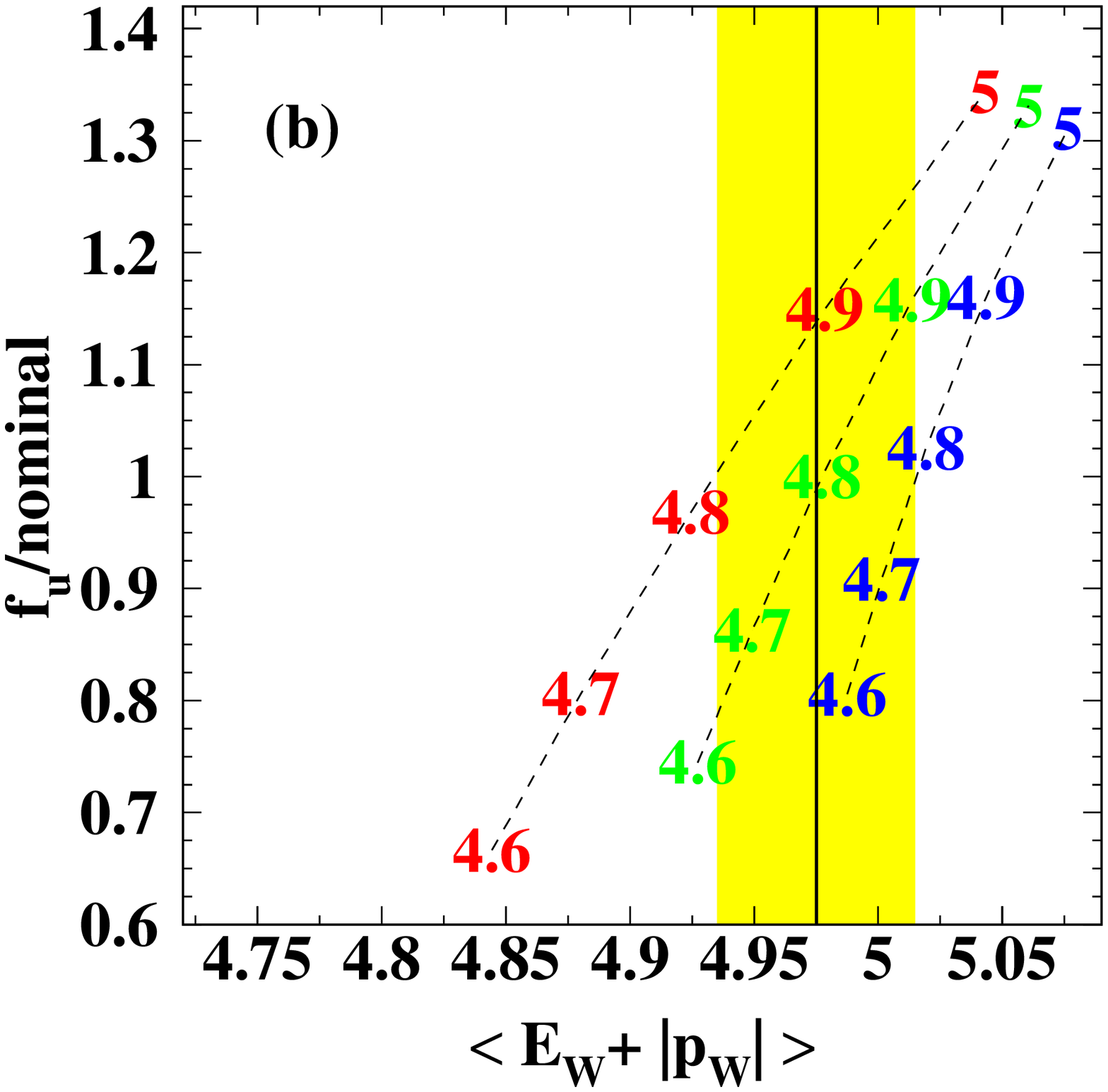}}\\
 \resizebox{0.49\textwidth}{!}{\includegraphics{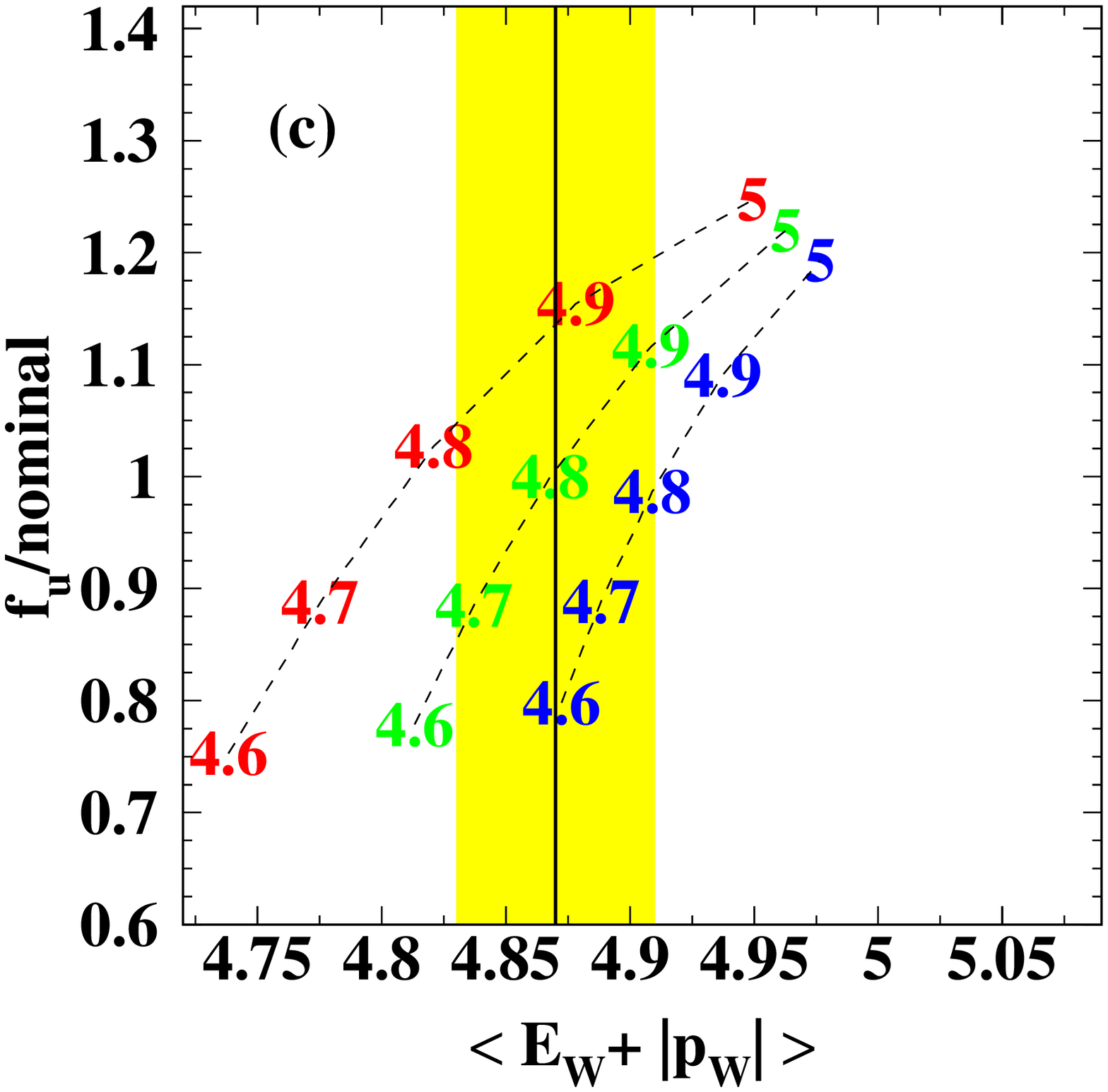}}
 \resizebox{0.49\textwidth}{!}{\includegraphics{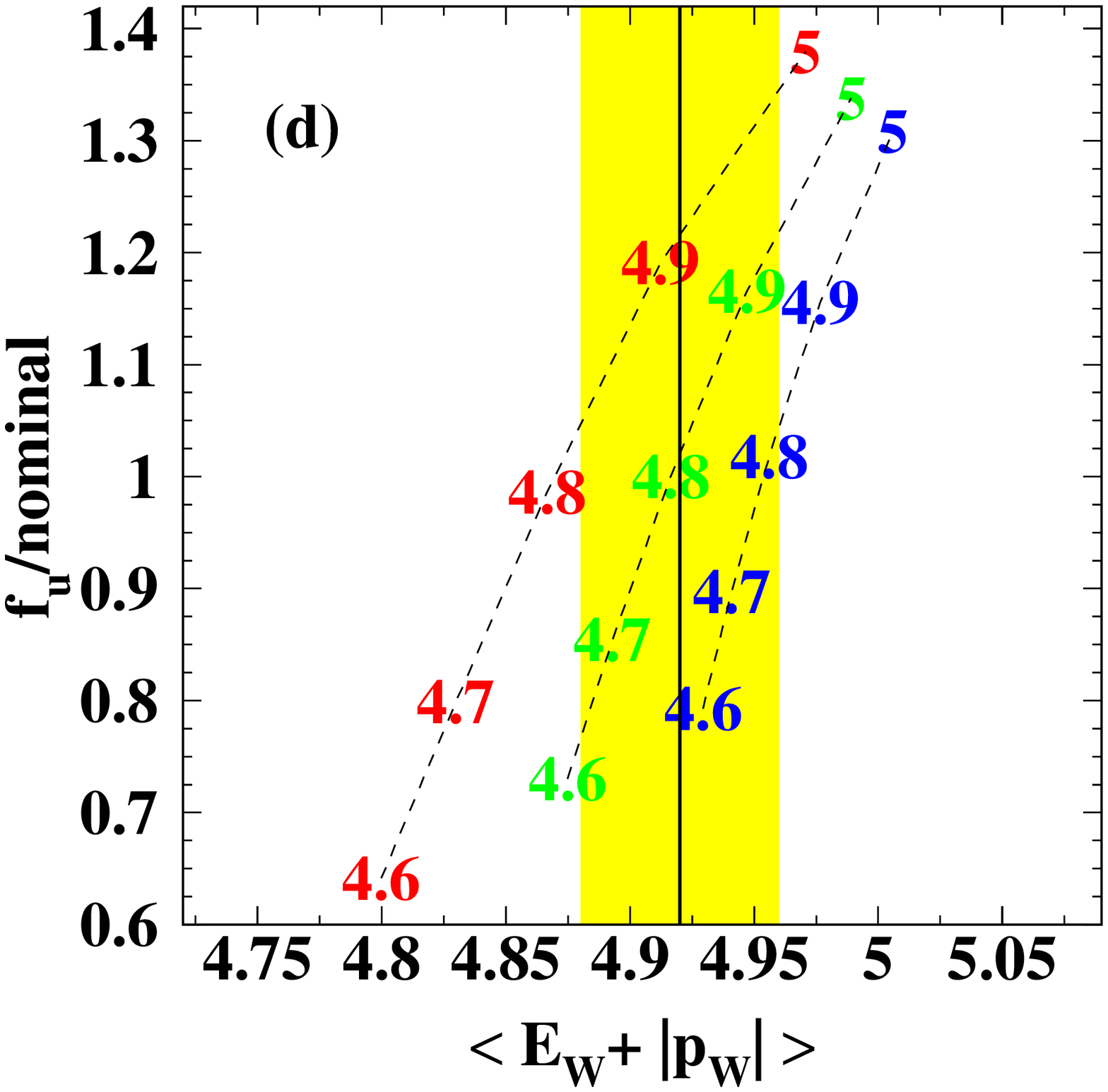}}
\end{center}
\begin{center}
\caption{The ratio $f_{\rm u}(m_{\rm b}, a)$/$f_{\rm u}(4.8, 1.29)$ is shown versus
$\EplusPave$ for different choices of the parameters 
$m_{\rm b}$ and $a$, with the $m_{\rm b}$ values indicated by the numbers on the
plot.
The leftmost, center and rightmost curves correspond to choices
$a=0.38, 1.29$ and $3.60$, respectively.  
The requirement $\shmax<3.2\,\GeV^2$
is used in (a);
in (b) the additional
requirement $E_\ell>2.1\,\GeV$ is made.
In (c) and (d) the requirements $\mH<1.7\,\GeV$ and $\mH<1.5\,\GeV$
are made, respectively.
In each case
the charged lepton energy is required to exceed $1\,\GeV$
in the ${\rm \FourS}$ frame.
The shaded bands show an example of
the impact a measurement of $\EplusPave$ with $0.04\,\GeV$ precision
would have. 
The values of $f_{\rm u}(4.8, 1.29)$ corresponding to requirements
(a)-(d) are, respectively, $0.22$, $0.15$, $0.67$ and $0.57$.
\label{fig-EplusP-fu}}
\end{center}
\end{figure}

The two parameters $m_{\rm b}$ and $a$ of the shape function
are varied to estimate the change in the fraction of
\bulnu\ decays in the signal region.  Fig.~\ref{fig-EplusP-fu}
shows the change in the
fraction $f_{\rm u}$ of \bulnu\ transitions in the signal region
versus $\EplusPave$
for different choices of shape function parameters.  
The variations $m_{\rm b} = (4.8 \pm 0.15)\,\GeV$
and $a =  1.29 {{+2.31}\atop{-0.91}}$ correspond to
$\overline{\Lambda} = (0.48 \pm 0.15)\,\GeV$ 
and $\lambda_1 = (-0.3 {{+0.15}\atop{-0.20}})\,{\GeV}^2$, respectively, 
and are a reasonable a priori choice at present.
The observable $\EplusPave$ has sensitivity to both $m_{\rm b}$ and $a$;
the sensitivity to $m_{\rm b}$ is reduced (i.e. 
$\Delta\EplusPave / \Delta m_{\rm b} < 1$) due to the
cuts used to reject $\bclnu$ decays.  

With no measurement
of $\EplusPave$ the uncertainty on $\Vub$ due to the shape function
(about $\sim 9\%$ for
the requirement $\shmax<3.2\,\GeV^2$)
is dominated by the uncertainty in $m_{\rm b}$.  A simultaneous measurement of
$\EplusPave$ could significantly reduce this uncertainty.  
For example, achieving a precision of $0.04\,\GeV$ on 
$\EplusPave$ reduces the uncertainty on $\Vub$
to $\sim 5\%$ for a cut on $\shmax<3.2\,\GeV^2$
(see the shaded band in Fig.~\ref{fig-EplusP-fu}).  
Similar results are obtained if $\mH<1.7\,\GeV$ is used
to select $\bulnu$ decays.  
If a stiff cut on the lepton energy is 
required in addition to the cut on $\shmax$, or if the cut on
$\mH$ is reduced to $1.5\,\GeV$, the theoretical
error is increased and the ability to reduce it with a given precision
on $\EplusPave$ is decreased.  However, a measurement
of $\EplusPave$ would 
provide an important cross-check on the externally chosen
parameter values in all cases.

\section{Discussion}

The use of $\EplusPave$ to constrain the shape function is analogous
to the use of the photon energy spectrum from \btosgam\ decays and
amounts to using the virtual ${\rm W}$ to probe the ${\rm b}$ quark
decay.  

In order to use the \btosgam\ spectrum 
to constrain the uncertainty on
\Vub, the $\gamma$ spectrum must be measured to as low an
energy as possible and deconvoluted to obtain an
estimate of $\overline{\Lambda}$ (and, perhaps eventually, 
$\lambda_1$).  Care 
must be taken to ensure that the theoretical expressions used
to extract $\overline{\Lambda}$ from \btosgam\ and to determine
the dependence of $\Vub$ on $\overline{\Lambda}$ are compatible.
Corrections to the universality of the shape function\cite{bsgamNow} 
must also be considered when applying this information to semileptonic
$\bulnu$ decays.

There are several advantages to the approach described in
this paper.  Since the same event generator is used for
extracting \Vub\ and determining the relationship between $\EplusPave$
and the shape function, no deconvolution is necessary.  The need for
a common theoretical framework for extracting and using the shape 
function parameters is trivially met; the parameters of the shape
function need not even be extracted.

Quantifying the shape function
parameters is of interest in its own right.  
In the absence of experimental cuts, 
the perturbative relationship between $\overline{\Lambda}$ and 
$\langle E_{\rm W}+|\vec{p}_{\rm W}|\rangle$ 
in the B meson rest frame
can be determined (to $\Order(\alpha_s)$) 
from formulae in Ref.~\cite{art:NeubertFazio}:
\begin{equation}
\overline{\Lambda} = m_{\rm B} - 
\langle E_{\rm W}+|\vec{p}_{\rm W}|\rangle -
\frac{91}{900}\frac{\alphas}{\pi}
\ \frac{2\, m_{\rm B}^2 - (2\, m_{\rm B} - 
\langle E_{\rm W}+|\vec{p}_{\rm W}|\rangle )^2}
{\langle |\vec{p}_{\rm W}|\rangle}\ .
\end{equation}
The relationship between $\overline{\Lambda}$ and the measured
$\EplusPave$ needs to be calculated in the presence of the
experimental cuts required to reject $\bclnu$ events.
Extracting $\lambda_1$ 
is not feasible due to the large intrinsic width
of the distribution and direct sensitivity to the
missing energy resolution.

The method outlined in this paper holds promise for reducing the
theoretical uncertainty in determining the fraction of $\bulnu$ decays
in the experimentally accessible region.
The \shmax\ variable
offers a way of extracting a clean signal for \bulnu\ decays with
modest theoretical uncertainties.  While it has a lower acceptance for
\bulnu\ decays than does the recoil mass $\mH$, it may be
advantageous experimentally because it does not require full
reconstruction of the second ${\rm B}$ meson in the event.  The use of
$\EplusPave$ may reduce the range of parameter space that need be
considered in evaluating the uncertainty on $\Vub$, and provides an
important cross-check on externally motivated parameter choices.
The ultimate experimental precision on $\Vub$ 
is hard to predict at present, but a precision approaching 5\%\ can be
envisaged.  There is strong
motivation to pursue such a measurement.


%
%

\end{document}